\documentstyle[12pt,epsf]{article}

\def\simge{\mathrel{\rlap{\raise 0.511ex \hbox{$>$}}{\lower 0.511ex 
\hbox{$\sim$}}}}
\def\simle{\mathrel{\rlap{\raise 0.511ex \hbox{$<$}}{\lower 0.511ex 
\hbox{$\sim$}}}} 
\def\slash#1{\setbox0=\hbox{$#1$}\dimen0=\wd0                     
\setbox1=\hbox{/} \dimen1=\wd1 
\ifdim\dimen0>\dimen1                    \rlap{\hbox to 
\dimen0{\hfil/\hfil}} #1                        
\else                                       
      \rlap{\hbox to \dimen1{\hfil$#1$\hfil}}  
      /   \fi}                                         
\newcommand{\be}{\begin{equation}}
\newcommand{\ee}{\end{equation}}
\newcommand{\bea}{\begin{eqnarray}}
\newcommand{\eea}{\end{eqnarray}}

\newcommand{\epse}{\varepsilon^{\prime}/\varepsilon}

\newcommand{\PL}[3]{{\it Phys.\ Lett.\ }         {\bf #1}, {#3} {(#2)}}

\newcommand{\PRL}[3]{{\it Phys.\ Rev.\ Lett.\ }  {\bf #1}, {#3} {(#2)}}
\newcommand{\PR}[3]{{\it Phys.\ Rev.\ }          {\bf #1}, {#3} {(#2)}}
\newcommand{\NP}[3]{{\it Nucl.\ Phys.\ }         {\bf #1}, {#3} {(#2)}}
\newcommand{\NPPC}[3]{{\it Nucl.\ Phys.\ (Proc. Suppl.) }         {\bf #1}, {#3} {(#2)}}

\newcommand{\JHEP}[3]{{\it JHEP\ }               {\bf #1}, {#3} {(#2)}}

\begin{document}

\begin{titlepage}
\begin{flushright}
RM3-TH/00-17 \\
ROMA-1297/00
\end{flushright}
\vskip 1. cm
\begin{center}
{\LARGE \bf   First Lattice Calculation of the 
\vskip0.3cm  Electromagnetic Operator Amplitude
\vskip0.3cm  $\langle \pi^0 \vert Q_\gamma^+ \vert K^0 \rangle$}
\vskip 1.4cm 
{\bf {\LARGE GLADIATOR}}
\vskip 0.2 cm 
{\bf {\LARGE The SPQ}cd{\LARGE R Collaboration}}
\vskip 0.2 cm
{\bf \large Southampton-Paris-Rome}
\vskip 1.0  cm
{\large\bf D.~Becirevic$^a$, V.~Lubicz$^b$, G.~Martinelli$^{a}$ and 
F.~Mescia$^a$}\\
\vspace{0.8cm}
{\normalsize {\sl 
$^a$ Dip. di Fisica, Univ. di Roma ``La Sapienza" and INFN,
Sezione di Roma,\\ Piazzale Aldo Moro 2, I-00185 Rome, Italy. \\
\vspace{.25cm}
$^b$ Dip. di Fisica, Univ. di Roma Tre and INFN,
Sezione di Roma Tre, \\
Via della Vasca Navale 84, I-00146 Rome, Italy. \\ \vspace{.25cm}
}
}
\vspace{.25cm}
\vskip1.cm
{\large\bf Abstract:\\[10pt]} \parbox[t]{\textwidth}{ 
We present the first lattice calculation of 
the matrix element of the electromagnetic operator
 $\langle \pi^0 \vert Q_\gamma^+ \vert K^0 \rangle$, where $Q_\gamma^+=
 (Q_{d}e/16 \pi^2)
        \left( {\bar s}_L \sigma^{\mu \nu} F_{\mu\nu} d_R +
               {\bar s}_R \sigma^{\mu \nu}  F_{\mu\nu} d_L 
               \right)$.  This matrix element plays an important r\^ole, since 
it contributes to  enhance  the CP
violating part of the  $K_L \to \pi^0 e^+ e^-$ amplitude 
in supersymmetric extensions of the  Standard Model. 	         
}  
\end{center}
\vspace*{1.cm} 

\end{titlepage}

\section{Introduction}
The origin of CP violation is one of the fundamental  questions of particle physics
and cosmology which  remains an open problem to date. 
The recent measurements of $\epse$~\cite{epsp99}
have definitively established direct CP violation and   ruled out superweak scenarios.
Unfortunately,  we are still far from a full quantitative description
of the  dynamics which generate the  amount of CP violation observed in 
 hadronic processes~\cite{epspth}.
Given the large theoretical uncertainties  affecting the calculation of $\epse$, it is
very useful to collect additional experimental information 
about CP violation in different processes.   The most interesting 
ones are those  for which CP violating effects are suppressed in the 
Standard Model (SM) and  enhanced in its extensions. 
Among the processes which have been considered in the literature,
we like to mention charge asymmetries in non-leptonic decays~\cite{Avilez,im}
and  CP asymmetries of hyperon decays~\cite{mura2}. 
\par 
Good candidates to provide new large CP violating effects 
are the supersymmetric extensions of the SM with generic 
flavour couplings and minimal particle content. In this framework,
among the possible contributions,
it has been recently recognized the importance of the 
electromagnetic  and chromomagnetic operators (EMO and CMO)
\bea
Q^\pm_\gamma &=& \frac{Q_{d}e}{16 \pi^2}
        \left( {\bar s}_L \sigma^{\mu \nu} F_{\mu\nu} d_R \pm
               {\bar s}_R \sigma^{\mu \nu}  F_{\mu\nu} d_L 
               \right) \label{eq:EMO} \\
Q^\pm_g &=& \frac{g}{16 \pi^2}
        \left( {\bar s}_L \sigma^{\mu \nu} t^a G^a_{\mu\nu} d_R \pm
               {\bar s}_R \sigma^{\mu \nu} t^a G^a_{\mu\nu} d_L 
               \right) \label{eq:CMO} 
\eea
The same mechanism, the  misalignment between quark and squark mass 
matrices, may indeed substantially increase  their CP-odd contribution  to 
physical processes. In previous studies, particular attention has been
devoted to the CMO which,  without conflict with the experimental
determination of the $K^0$--$\bar K^0$ mixing amplitude,
 can account for the largest part of the measured 
 $\epse$~\cite{MM}--\cite{BCIRS}.
\par In this paper, we consider the CP violating contribution of the 
EMO to $K_L \to \pi^0 e^+ e^-$.   The master formula which 
has been used in the  numerical calculation of the rate is~\cite{im}
\be 
{\rm B}(K_L \to \pi^0 e^+ e^-)_{EMO} =
5.3 \times 10^{-4} 
\left(\frac{\tilde y_\gamma(m_{\tilde g}, x_{gq}) G_0(x_{gq})}
{\tilde y_\gamma(500 {\rm GeV}, 1) G_0(1)}\right)^2 
\tilde B_T^2 \ (\mbox{Im} \delta_+)^2 \, . \label{eq:nschifo}
\ee
The coupling $\delta_+$ is related to the  splitting
in the down-type squark mass matrix. 
The definitions of $\delta_+$  and $\tilde y_\gamma$ can be found 
in sec.~\ref{sec:uno} and  
$x_{g q}=m^2_{\tilde g}/m^2_{\tilde q}$ is the ratio of 
gluino and (average) squark mass squared. The numerical coefficient in
eq.~(\ref{eq:nschifo}) is the appropriate one for the operator $Q_\gamma^+$
renormalized in $\overline{MS}$ at the scale $\mu=2$~GeV.
\par Our main result is  the first lattice calculation of the
$B$ parameter $\tilde B_T$ (also defined in sec.~\ref{sec:uno}),
for which we obtain
\be \tilde B^{\overline{MS}}_T(\mu= 2
{\rm GeV}) = 1.21  \pm  0.09 \pm 0.04^{+0.07}_{-0.00} \ , \label{eq:bt}  \ee
where the first error is the statistical one, the second is the systematic 
error due to the uncertainty on the ratio of the EMO  to
the vector current matrix elements and the third is the error coming from the
uncertainty on the renormalization of the magnetic operator.  
\par  From the experimental upper bound~\cite{rari} 
\bea 
 {\rm B}(K_L \to \pi^0 e^+ e^-) < 5.1 \times 10^{-10}
\label{eq:exps} \,  
\eea
by  taking $\tilde B_T$ from
eq.~(\ref{eq:bt}), $x_{gq }=1$ and  $m_{\tilde g}=500$~GeV,
and using   eq.~(\ref{eq:nschifo}),   we obtain 
\be \vert \mbox{Im} \delta_+ \vert  < 1.0 \times 10^{-3} \quad 
\quad  (95\, \% \ C.L.) \ee
\par
The remainder of the paper is organized as follows: in sec.~\ref{sec:uno} 
 all the formulae necessary to derive eq.~(\ref{eq:nschifo}) are presented;
in sec.~\ref{sec:due} we describe the lattice simulation  
and discuss the calculation of  $\tilde B_T$;
 sec.~\ref{sec:conclusion} contains our conclusion.
\section{EMO Contribution to $ {\rm B}(K_L \to \pi^0 e^+ e^-)$} 
\label{sec:uno}
In this section, we recall the main ingredients necessary to compute
$ {\rm B}(K_L \to \pi^0 e^+ e^-)$ in SUSY  and introduce all 
the quantities appearing in eq.~(\ref{eq:nschifo}). 
We discuss separately the effective Hamiltonian and the calculation of
the branching ratio.
\subsection{The effective Hamiltonian for the magnetic operators}
\label{subsec:eh}
\par 
The supersymmetric contribution to
the effective Hamiltonian,  in the case of the magnetic operators, 
 can be written as
\bea {\cal H}_{MO}= C_\gamma^+(\mu) Q_\gamma^+(\mu) +  C_\gamma^-(\mu)
 Q_\gamma^-(\mu)
+ C_g^+(\mu) Q_g^+(\mu) +  C_g^-(\mu) Q_g^-(\mu) \label{eq:eh}
\ , \eea
where the operators are renormalized at the scale $\mu$.
The Wilson coefficients  generated by gluino exchanges 
at the SUSY breaking scale are given by~\cite{BCIRS,GGMS} 
\bea  C_\gamma^\pm(m_{\tilde g})&=&
 \frac{\pi \alpha_s(m_{\tilde g})}{m_{\tilde g}}
 \left[ \left(\delta^{D}_{LR}\right)_{21} \pm
   \left(\delta^{D}_{LR}\right)^*_{12} \right] F_0(x_{g q})\,  , \nonumber  \\
C^\pm_g(m_{\tilde g}) &=& \frac{\pi \alpha_s(m_{\tilde g})}{m_{\tilde g}}
 \left[ \left(\delta^{D}_{LR}\right)_{21} \pm
   \left(\delta^{D}_{LR}\right)^*_{12} \right] G_0(x_{g q})\, 
\eea
Here $(\delta^D_{LR})_{ij}=(M^2_D)_{i_L j_R}/m^2_{\tilde q}$ 
denote the off-diagonal entries of the (down-type) squark mass 
matrix in the super-CKM basis \cite{HKR}.
The explicit expressions of  $F_0(x)$ and  $G_0(x)$
are:
\bea F_0(x)&=& \frac{4 x \left(1 + 4 x - 5 x^2 + 4 x \ln\left(x\right)
 + 2 x^2 \ln\left(x\right) \right)}{3 (1-x)^4}    \label{eq:f0} \, , \\
 \nonumber \\
 G_0(x)&=& \frac{ x \left(22 -20 x - 2 x^2 + 16 x \ln\left(x\right)
 - x^2 \ln\left(x\right) + 9 \ln\left(x\right)\right)}{3 (1-x)^4}  \ ,   
 \label{eq:g0}  \eea
 with $F_0(1)=2/9$ and $G_0(1)=-5/18$.
In the following we will use the combinations
$\delta_\pm = (\delta_{LR}^D)_{21} \pm (\delta_{LR}^D)^*_{12} = 
(\delta_{LR}^D)_{21} \pm (\delta_{RL}^D)_{21}$. 
These quantities are the  natural 
couplings appearing at first order in any  parity conserving ($+$) or
parity violating ($-$) observable.
\par In the $( Q_\gamma^\pm, Q_g^\pm )$ basis, using the leading order (LO) anomalous
dimension matrix 
\bea
&& \hat \gamma \ = \ 
 \  
 \left( \begin{array}{cc}  
 \frac{8}{3} 
 &\; 0
\\
 
 &\; 
\\
{{32}\over {3}} &\;  
{{4}\over {3}}
\\
\end{array} \right) \  , \eea
it is straightforward to derive
\bea 
 C_\gamma^\pm(\mu)&=& \eta ^2 \left[  C_\gamma^\pm(m_{\tilde g}) +
 8 (1-\eta^{-1} )  C_g^\pm(m_{\tilde g})\right] \ , \nonumber \\
 \nonumber \\
 C^\pm_g(\mu) &=& \eta \, C_g^\pm(m_{\tilde g}) \, , \label{eq:coefs} \eea
 where 
\bea \eta = \left( \frac{\alpha_s(m_{\tilde g})}{\alpha_s(m_t)} \right)^{2/21}
\left( \frac{\alpha_s(m_t)}{\alpha_s(m_b)} \right)^{2/23}
\left( \frac{\alpha_s(m_b)}{\alpha_s(\mu)} \right)^{2/25} \ . \label{eq:eta}
 \eea
\subsection{Calculation of $ {\rm B}(K_L \to \pi^0 e^+ e^-)$}
In order to compute the rate, besides the Wilson coefficient,
we also need the matrix element of the operator $Q^+_\gamma$,
which is usually expressed in term of a suitable $B$ parameter~\cite{BCIRS}
\be \langle \pi^0 \vert Q^+_\gamma \vert K^0 \rangle
= i \frac {Q_d e  \sqrt{2}}{16 \pi^2 m_K} p_\pi^\mu p_K^\nu F_{\mu\nu} B_T
{\cal R}_T(q^2)\, . \label{eq:btdef} \ee
With respect to the standard definition of $B_T$,  we have introduced 
the $q^2$-dependent form factor ${\cal R}_T(q^2)$ (${\cal R}_T(0)=1$)
to account for the dependence of the matrix element 
on the momentum transfer $q=p_K-p_\pi$. Note that, since we are using 
renormalized operators,  $B_T$ depends
 on both the renormalization scheme and scale.
\par Neglecting  lepton masses,  and isospin breaking effects,
we may write the following useful identity
\be  \frac{\langle \pi^0 e^+ e^-\vert Q^+_\gamma \vert K^0 \rangle}
{{\cal R}_T(q^2)}= 
\frac{Q_d \alpha B_T }{4 \pi m_K  f^+(q^2)}
 \langle \pi^0 e^+ \nu_e \vert (\bar \nu_e \gamma_\mu e) (\bar s \gamma^\mu
 u)  \vert K^+ \rangle \, , \label{eq:rel}\ee
where   $\alpha$ is the electromagnetic coupling
and $f^+(q^2)=f^+(0) {\cal R}_+(q^2)$ is the vector current form factor
defined as 
\be   \langle \pi^0  \vert \bar (\bar s \gamma^\mu
 u)  \vert K^+ \rangle = {1\over \sqrt{2}} \left[
\left(p_K^\mu + p_\pi^\mu -\frac{m_K^2-m_\pi^2}{q^2}
q^\mu \right) f^+(q^2)  +  \frac{m_K^2-m_\pi^2}{q^2} q^\mu  f^0(q^2)\right] \ .  
\label{eq:fpf0} \ee
Eq.~(\ref{eq:rel}) allows us to write 
${\rm B}(K_L \to \pi^0 e^+ e^-)_{EMO}$ in terms of the $K^+$
semileptonic  branching ratio
\be  {\rm B}(K_L \to \pi^0 e^+ e^-)_{EMO}= 2 \left(\frac{\alpha}{2 \pi}\right)^2
{\rm B}(K^+\to \pi^0 e^+ \nu_e) \frac{\tau(K_L)}{\tau(K^+)}
\frac{\vert {\rm Im} \Lambda^+_g \tilde y_\gamma\vert^2}{\vert V_{us} \vert^2}
\tilde B_T^2  \, , \label{eq:allows} \ee
where we have followed the  notation of ref.~\cite{BCIRS} by defining
\be    \mbox{Im} \Lambda^+_g \tilde y_\gamma= 
\frac{Q_d}{\sqrt{2} G_F m_K} \mbox{Im} C^+_\gamma 
\ , \ee
with 
\bea \Lambda^+_g (x_{gq})&=& \delta _+ \ G_0(x_{gq}) \nonumber \\
\tilde y_\gamma(m_{\tilde g},x_{gq}) &=&  \frac{\pi \alpha_s(m_{\tilde g})}{m_{\tilde g}}
\frac{Q_d}{\sqrt{2} G_F m_K}  \eta^2 \left[ \frac{F_0(x_{gq})}{G_0(x_{gq})}
+ 8 \left(1-\eta^{-1}\right)\right] \ . \label{eq:defdef} \eea
The definition of $\tilde y_\gamma$ given above differs from that of 
ref.~\cite{BCIRS} by a factor $-B_T$, since it is preferable to separate the 
Wilson coefficient from the $B$ parameter.
\par We have introduced the ``effective" $B$ parameter $\tilde B_T$ defined
as 
\be \tilde B_T = \frac{B_T}{f^+(0)} \times 
T \ , \quad \quad   T= \frac{\int_0^{(m_K-m_\pi)^2} \, dq^2
\lambda^{(3/2)}(q^2) \vert{\cal R}_T(q^2)\vert^2}
{\int_0^{(m_K-m_\pi)^2 }\, dq^2
\lambda^{(3/2)}(q^2) \vert {\cal R}_+(q^2)\vert^2 }\ , \ee
where 
$\lambda(q^2)=  (m_K^2+m_\pi^2-q^2)^2-4 m_K^2 m_\pi^2$.
$T$ is the correction due to the different $q^2$ dependence
of the tensor and vector form factors.  In the calculation of
$T$, we have used the experimental determination of the $q^2$ dependence
of the semileptonic decay rate~\cite{lpiu}.
\be {\cal R}_+(q^2) = 1 + \frac{\lambda_+}{m_\pi^2} q^2
\, , \quad \quad \lambda_+ = 0.0286 \pm 0.0022 \label{eq:l+} \ee
and the slope of the tensor form factor extracted from our lattice
data (see sec.~\ref{sec:due}) 
\bea {\cal R}_T(q^2) =  1 + \frac{\lambda_T}{m_\pi^2}   q^2 \ , \quad \quad  
\lambda_T = 
0.022\pm 0.001 \, . \label{eq:lt} \eea
Since the correcting factor is very close to one, $T=0.99$,  
its effect is practically
negligible for the value of the effective $B$ parameter, $\tilde B_T$.
For the same reason, the difference between  our value of
$\lambda_+$ in eq.~(\ref{eq:lt}) and the experimental value 
of eq.~(\ref{eq:l+}) is ininfluential to the determination of
$\tilde B_T$.
\par
In order to compute $\tilde B_T$,  we also need $B_T$ and $f^+(0)$. 
To wit, using the data discussed in sec.~\ref{sec:due}, we have 
followed  two different  procedures:
\begin{itemize}
\item  we have taken the value  $f^+(0)=0.978$ from ref.~\cite{gl}.
This number was obtained by neglecting  isospin breaking effects.
 Using our result for the $B$ parameter, $B_T= 1.23 \pm 0.09$, and
$T=0.99$, we  get $\tilde B_T= 1.25 \pm 0.09$;
\item we have computed on our data 
the ratio $B_T/f^+(0)$ extrapolated to the physical  meson masses,
obtaining $B_T/f^+(0)=1.18 \pm 0.09$.
With the same value of $T$ as before, in this case we get 
  $\tilde B_T= 1.17 \pm 0.09$.
\end{itemize} 
The difference between the two different procedures is taken into 
account in the 
 systematic error, so that we get
\be \tilde B_T =1.21  \pm  0.09 \pm 0.04 \ .\label{eq:final} \ee 
To this result, we add a very generous estimate ($+ 6\%$) of the 
systematic error (to be discussed in the next section)
due to the renormalization of the EMO. In this way we arrive to the result
quoted in eq.~(\ref{eq:bt})  of the introduction.
This result is consistent with previous  estimates from
refs.~\cite{ria,cip}.
\par We have now all the necessary elements for the calculation of 
${\rm B}(K_L \to \pi^0 e^+ e^-)_{EMO}$. Using eq.~(\ref{eq:allows}), 
the definitions (\ref{eq:defdef}) and the values given in 
table~\ref{tab:parameters}, we arrive to the master formula 
in eq.~(\ref{eq:nschifo}) which has been used, together with $\tilde B_T$,
to constrain ${\rm Im} \delta_+$.   
\begin{table}
\vspace*{-2cm}
\begin{center}
\begin{tabular}{|c|c|} 
\hline
{\phantom{\Large{l}}}\raisebox{+.1cm}{\phantom{\Large{j}}}
{\sl Parameter} & {\sl Value and error} \\ \hline
$G_F$ & $1.16639\times 10^{-5} \mbox{GeV}^{-2}$ \\
$V_{us}$ & $0.2196 \pm 0.0023$ \\
$m_K$ & $0.498$ GeV\\
$m_\pi$ & $0.135$ GeV\\
$\tau(K_L)$ & $(5.15 \pm 0.04) \times 10^{-8}$ s \\
$\tau(K^+)$ & $(1.2385 \pm 0.0025) \times 10^{-8}$ s\\ 
${\rm B}(K^+\to \pi^0 e^+ \nu_e)$ & $0.0485\pm 0.0009$ \\
$\alpha_s(M_Z)$ & $0.119 \pm 0.003$ \\
 \hline 
\end{tabular}
\caption[]{\label{tab:parameters}{\sl Average and errors of 
the main parameters.
When the error is negligible it has been omitted.}}
\end{center}
\end{table}  
\section{Lattice calculation of the EMO matrix elements}
\label{sec:due}
In this section, we describe the procedure followed to obtain, in  our lattice simulation, the 
$B$ parameter, $B_T$ ($f^+(0)$), and the form factor, ${\cal R}_T(q^2)$ 
(${\cal R}_+(q^2)$),
necessary to the  computation  of ${\rm B}(K_L \to \pi^0 e^+ e^-)_{EMO}$.
Since the calculation of the form factors on the lattice  has been discussed in 
several papers, see for example \cite{oursl} and references therein,
we only give here  the details which characterize the present  study.
\par All our lattice results have been obtained using a non-perturbatively improved 
action~\cite{luscher}. The relevant operators, namely the vector and 
tensor currents
\bea \hat V_\mu &=& Z_V (1 + b_V m a) \left[  \bar q \gamma_\mu Q + 
i c_V \partial_\nu \bar q \sigma_{\mu\nu} Q \right] \ , \nonumber \\
\nonumber \\
\hat T_{\mu\nu} &=& Z_T(\mu)  (1 + b_T m a) \left[  i \bar q \sigma_{\mu\nu}  Q + 
c_T \left(\partial_\nu \bar q \gamma_{\mu} Q -\partial_\mu \bar q \gamma_{\nu} Q \right) \right] \ , 
\eea
are improved. In our study, the coefficient $b_T$,
computed   at ${\cal O}(\alpha_s)$~\cite{weisz}, 
is  evaluated by using  boosted  perturbation theory~\cite{lpm}.  
$Z_T(\mu)$  was obtained in ref.~\cite{leroy} 
with the non-perturbative method of ref.~\cite{np},
in the RI-MOM scheme at the renormalization scale $\mu$. Note that
for the tensor current,  at the NLO,
 the RI-MOM scheme in the Landau gauge coincides with
the $\overline{MS}$ scheme. 
The other constants are taken  from the most recent non-perturbative 
determinations~\cite{sisi,lnal}.  In summary,
we have used the following values
\bea &\,&Z_V = 0.79 \ , \  \quad b_V= 1.4   \ , \
  \quad c_V= -0.09 \ , \nonumber \\
&\,& Z_T(\mu=2\ \mbox{GeV}) = 0.87 \pm 0.01 \ , \
  \quad b_T= 1.2 \ , \  \quad c_T=0.05  \ . \eea
Since the perturbative value of $Z_T$ is $Z_T(\mu=2\ \mbox{GeV}) = 0.934(5)$,
 we estimate that the  systematic error due to the normalization of 
the lattice operator  is less than $6\, \%$. Also this effect is included in
the evaluation of the systematic error.
\par Our analysis is based on a sample of $91$ independent quenched gauge-field 
configurations, generated at the lattice coupling constant $\beta=6/g_0^2=6.2$, on the volume $24^3\times
48$. 
We use the combination of values of the hopping parameter, $K_{L,l}$,
which are given  in tab.~\ref{tab:1}.  In order to calibrate the lattice spacing, $a$, and to extrapolate
the form factors to the physical meson masses, we use the lattice 
plane method~\cite{lplme}.
We find $a^{-1}=2.7\pm 0.1$~GeV, in agreement with previous simulations.

\subsection{Extraction of the form factor}
Using standard lattice techniques, we have extracted from suitable correlation
functions the matrix element of the operator
\be \langle \pi^0 \vert \bar s \sigma^{\mu\nu} d\vert K^0 \rangle
= i \, \left( p_K^\mu p_\pi^\nu -p_K^\nu p_\pi^\mu \right) \frac{\sqrt{2}
 f_T(q^2)}{m_K+m_\pi} \label{eq:defm} \ , \ee
where we have introduced the form factor $f_T(q^2)$ to make contact 
with the definition used in
ref.~\cite{oursl}.  From eq.~(\ref{eq:btdef}), one finds
\be \frac{2   f_T(q^2)}{m_K+m_\pi}=\frac{B_T   {\cal R}_T(q^2)}{m_K}
\ , \ee
so that $f_T(0)=B_T$ for $m_K=m_\pi$ corresponding to degenerate quark masses,
$m_s=m_{u,d}$.
\par
\begin{figure}[h!]
\begin{center}
\begin{tabular}{c}
\epsfxsize16.0cm\epsffile{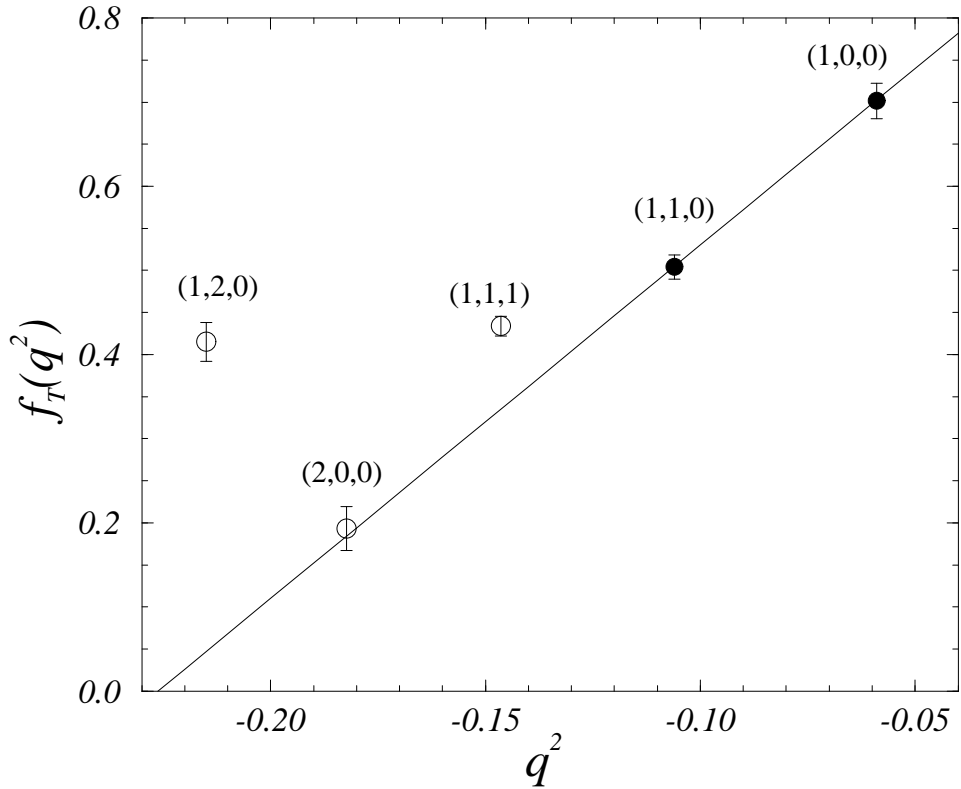} 
\end{tabular}
\caption{\sl $f_T(q^2)$ as a function of $q^2$ for 
$K_L=K_l=0.1344$. Beside every lattice point we
mark the components $(n_x,n_y,n_z)$ of the three momentum given to the pion in
lattice units ($2\pi/La$). 
\label{fig:ftq2} }
\end{center}
\end{figure}

 Besides $f_T(q^2)$,
we have also considered the vector form factors 
 $f^+(q^2)$ and $f^0(q^2)$ ($f^+(0)=
f^0(0)$) appearing in eq.~(\ref{eq:fpf0}). 
$f_T(q^2)$,   $f^+(q^2)$ and $f^0(q^2)$ have  been  computed for  $\vec p_K=0$
at  several values  of the pion  momentum,  
$\vec p_\pi=(2 \pi/L) (n_x,n_y,n_z)$ in lattice units. We have results for  
$(n_x,n_y,n_z)=(0,0,0)$, $(1,0,0)$, $(2,0,0)$, $(1,1,0)$, $(1,1,1)$
and $(1,2,0)$.   The results for  $f_T(q^2)$ as a function of
the dimensionless variable $q^2 a^2$ are shown in 
fig.~\ref{fig:ftq2}.
\par At fixed quark masses, we fit the lattice form factors 
 to the expressions
\bea
f_T(q^2) =  f_T(0)\left(1\,+\, \alpha_T \, q^2 a^2\right)\ ,  \quad \quad
f^+(q^2) =  f^+(0)\left(1\,+\, \alpha_+ \, q^2 a^2\right) \ .
\eea
The slopes in eqs.~(\ref{eq:l+}) and (\ref{eq:lt}) are given
by $\lambda_{+,T}= \alpha_{+,T} m^2_\pi a^2$. 
For $f_T(q^2)$ we have only used the points corresponding to  
$(1,0,0)$, $(1,1,0)$,  labeled as squares in 
 fig.~\ref{fig:ftq2}, 
because the quality of the signal in the other cases is rather poor,
and gets worse as the quark masses decrease.
For completeness, in the figure, we have also 
shown the  other points which have not been considered for the fit.
The results for  $f_T(q^2)$
at different values of the ``strange" and light 
quark masses, corresponding to the hopping parameters $K_L$ and $K_l$
respectively, are given in tab.~\ref{tab:1}. 
\begin{table}[h!]
\begin{center}
\begin{tabular}{||c|c|c|c|c|c||}
\hline\hline \multicolumn{6}{||c||}{$\langle \pi^0 \vert \bar s \sigma_{\mu\nu}
d \vert K^0\rangle$} \\ \hline
$K_L$ & $K_l$ & $f_T(0)$ &  $\alpha_T$    &  $M^2_{P}(K_L,K_l)$ & 
$M^2_{P}(K_l,K_l)$\\ \hline
0.1344  & 0.1344 &  0.95(4) & 4.42(23)& 0.090(1) & 0.090(1) \\
0.1344  & 0.1349 &  0.86(4) & 4.85(31)& 0.073(1) & 0.058(1)\\
0.1344  & 0.1352 &  0.90(5) & 6.48(24)& 0.064(1) & 0.039(1)\\
0.1349  & 0.1349 &  0.84(5) & 6.05(35)& 0.058(1) & 0.058(1)\\
0.1349  & 0.1352 &  0.81(5) & 6.86(29)& 0.049(1) & 0.039(1)\\
0.1352  & 0.1352 &  0.80(6) & 7.21(48)& 0.039(1) & 0.039(1)\\\hline\hline
\end{tabular}
\end{center}
\caption{{\sl $f_T(0)$  and the $\alpha_T$ parameter for different 
combinations
of the mesons masses, $M_P$, in lattice units.}}
\label{tab:1}
\end{table}
\subsection{Extrapolation to the physical point}
The values of $f_T(0)$ and $\alpha_T$ in tab.~\ref{tab:1}
have been extrapolated to the physical point, corresponding to
$M_P(K_L,K_l)=m_K a$ and $M_P(K_l,K_l)=m_\pi a$,
with the lattice-plane method of ref.~\cite{lplme}.  
Two different formulae have been used:
\begin{itemize}
\item  we have ignored the  SU(3) symmetry breaking  corrections, due to the 
$m_s$--$m_{u,d}$ mass 
difference,   by making a fit of the form 
\begin{equation}
y=C + L \, M^2_P(K_L,K_l) \ , 
\label{eq:su3-no}
\end{equation} 
where $y=f_T(0)$ or $\alpha_T$. The  results are
\begin{eqnarray}
f_T(0)&=&0.77(6)\;\;\;C_{f_T(0)} =0.67(8) \;\;\;L_{f_T(0)}= 2.9(6) \nonumber\\
 \\
\alpha_T&=&7.8(4)\;\;\;C_{\alpha_T} =9.9(6) \;\;\;L_{\alpha_T}=-60(7)
\nonumber 
\end{eqnarray}
\begin{figure}[h!]
\begin{center}
\begin{tabular}{c}
\epsfxsize16.0cm\epsffile{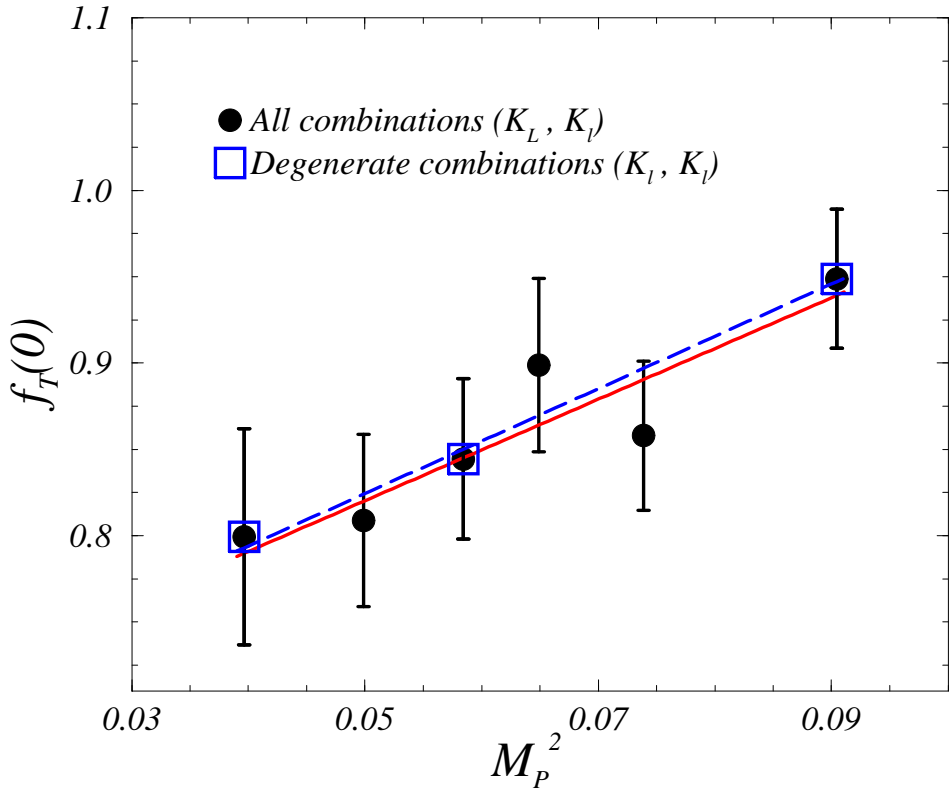} 
\end{tabular}
\caption{\label{fig:fq3}{\sl $f_T(0)$  as a function of the 
squared pseudoscalar meson 
mass in lattice units. The full (dashed) line represents a fit of 
the lattice points 
to eq.~(\ref{eq:su3-no}) for all (degenerate) meson masses.}}
\end{center}
\end{figure}
\begin{figure}[h!]
\begin{center}
\begin{tabular}{c}
\epsfxsize16.0cm\epsffile{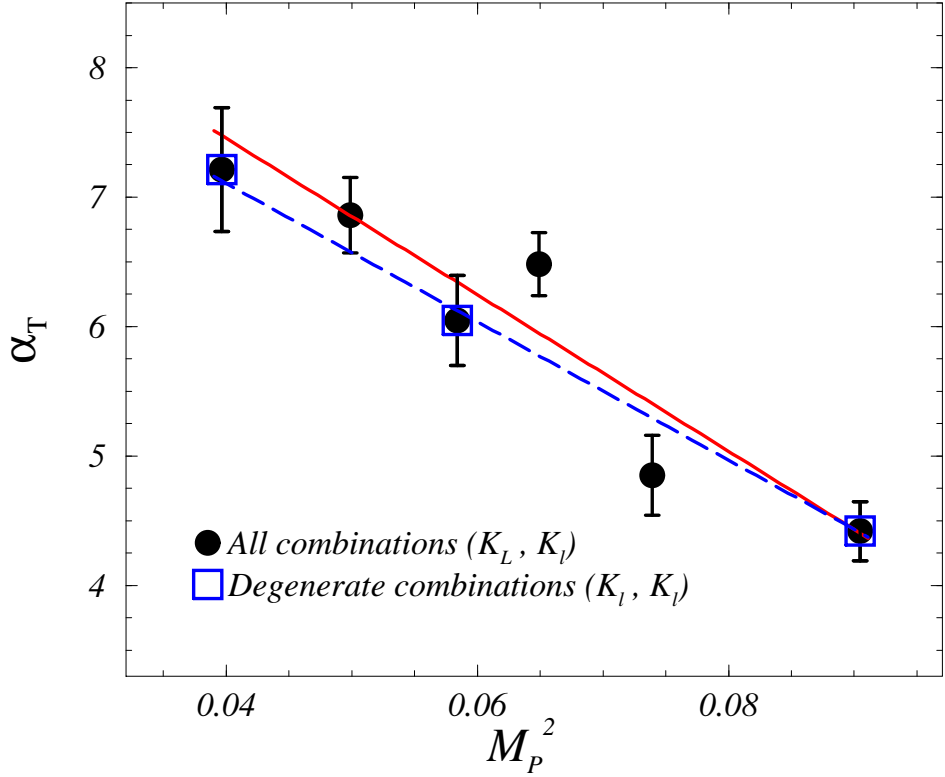}  
\end{tabular}
\caption{\label{fig:fq2}{\sl  $\alpha_T$ as a function of the squared 
pseudoscalar meson 
mass. The curves represent a fit of the lattice points to the 
eq.~(\ref{eq:su3-no}).}}
\end{center}
\end{figure}
From figs.~\ref{fig:fq3}~and~\ref{fig:fq2},
 we see  that SU(3) breaking effects are 
small since all the points are rather close to the straight lines of the fit.
The two lines correspond either to
 a fit to  all the  points or to those corresponding
to degenerate mesons ($M_P(K_L,K_l)=M_P(K_l,K_l)$) only.  
\item  we have chosen for $f_T(0)$ and $\alpha_T$  a fitting
formula which accounts for SU(3) breaking effects
\begin{equation}
y =C + \tilde L \, M^2_P(K_L,K_l) + L \, M^2_P(K_l,K_l) 
\label{eq:su3-ro}
\end{equation} 
In this case the results read 
\begin{eqnarray}
f_T(0)&=&0.78(6)\;\;\;C_{f_T(0)} =0.67(8) 
\;\;\;\tilde L_{f_T(0)}=3.1(7)\;\;\;L_{f_T(0)}=-0.14(72) \nonumber \\
\label{eq:ftfinal}\\  
\alpha_T &=&8.1(4)\;\;\;C_{\alpha_T} =9.5(7) \;\;\;\tilde L_{\alpha_T}
=-38(10)\;\;\;L_{\alpha_T}=-20(4) \nonumber 
\end{eqnarray}
We conclude that SU(3) breaking  effects are very  small  for
$f_T(0)$ and quite small for $\alpha_T$.  
In the following, for physical applications we will use the results
in eq.~(\ref{eq:ftfinal}).
\end{itemize}
Using the physical $K^0$ and $\pi$ masses, from  eq.~(\ref{eq:ftfinal})
we obtain 
\bea B_T = 1.23 \pm 0.09 \, , \quad \quad \lambda_T= 0.022 \pm 0.001 \ , 
\label{eq:primo}\eea
at a renormalization scale $\mu=2$~GeV.  If one needs the operator
at  another value of $\mu$, the slope will remain the same, whereas
$B_T$ scales according to the formula~\cite{Gracey}
\bea B_T(\mu_2) &=& 
\left(\frac{\alpha_s(\mu_2)}{\alpha_s(\mu_1)} \right)^{4/(33-2\, n_f)}
\left[ 1 \ +\ \frac{2}{9}\left({12411-126\, n_f +52\,  n_f^2 \over (33 - 2 \, n_f)^2}\right) 
{\alpha_s(\mu_2)-\alpha_s(\mu_1)\over 4 \pi}
 \right]
B_T(\mu_1)  \nonumber \\
 &=& 
\left(\frac{\alpha_s(\mu_2)}{\alpha_s(\mu_1)} \right)^{4/33}
\left[ 1 \ +\ {1379\over 2178} {\alpha_s(\mu_2)-\alpha_s(\mu_1)\over \pi}
 \right]
B_T(\mu_1) \, ,\label{eq:evol} \eea
where we set $n_f=0$, since we are working in the quenched approximation. In ref.~\cite{leroy} 
it was shown that even the one-loop evolution 
gives a very satisfactory description 
of the scale dependence of  the matrix elements of
the non-perturbatively renormalized EMO.
\par
We also present our  result for the ratio $B_T/f^+(0)$ extrapolated to the 
physical point
\be \frac{B_T}{f^+(0)}= 1.18 \pm 0.09 \  .\label{eq:secondo} \ee
This number is lower than the result obtained by combining $B_T=1.23 \pm 0.09$
and $f^+(0)=0.978$, namely $B_T/f^+(0)=1.26 \pm 0.09$. This happens because,
on our data, we get $f^+(0)=1.04 \pm 0.06$ , which is $6\%$ larger than the
result of ref.~\cite{gl}.   
 This is why it is 
important to have a direct determination of   $B_T/f^+(0)$:
 systematic effects are expected to be smaller  in the ratio. Moreover,
the comparison between the two ways to compute $B_T/f^+(0)$ allows us
to evaluate  the  systematic uncertainty.
Indeed the difference between the two way of determining 
$B_T/f^+(0)$ is compatible in size with the uncertainty which can
be estimated by  extrapolating the data using different procedures. In this case,
the different extrapolations
 give results  varying by about $5 \%$.  Thus we take $5 \%$
as a measure of a further systematic uncertainty on the determination of the ratio
$B_T/f^+(0)$.
From eqs.~(\ref{eq:primo}), using 
$f^+(0)=0.978$,   and  from eq.~(\ref{eq:secondo}), we obtain 
$\tilde B_T= 1.25\pm 0.09 $ and $\tilde B_T = 1.17\pm 0.09$, respectively.
Considering the difference as a systematic error,
we end up with our final result, which was given already in
eq.~(\ref{eq:final}). 

\section{Conclusion}
\label{sec:conclusion}
We have presented the first lattice calculation of the matrix element
 $\langle \pi^0 \vert Q_\gamma^+ \vert K^0 \rangle$. The operator is 
renormalized non-perturbatively in the RI-MOM scheme  which,  at the NLO,
is equal to the $\overline{MS}$ scheme.  Including the statistical and systematic 
uncertainties (except  quenching),  our final result has an error smaller than 
$10 \%$.  This allows,  from the 
upper limit on ${\rm B}(K_L \to \pi^0 e^+ e^-)$,   to put more stringent bounds on
 squark-mass differences in supersymmetric extensions of the Standard Model.
\newpage
\section*{Acknowledgements}
We   thank G.~Isidori for  illuminating discussions. 
G.M. acknowledges  the warm and stimulating hospitality of the LPT du Centre
d'Orsay where a large  part of this project has been developed.
We acknowledge the M.U.R.S.T. and the INFN for partial support.

\end{document}